\documentclass[twocolumn]{aastex62}

\usepackage{hyperref}
\usepackage{amsmath}

\received{January 1, 2018}
\revised{January 7, 2018}
\accepted{\today}
\submitjournal{ApJ}

\shorttitle{Gravitational Wave Sources from Dwarf Galaxy Mergers}
\shortauthors{Conselice et al.}

\def\solm{M$_{\odot}\,$}

\def\kms{km s$^{-1}$}

\def\solm{M$_{\odot}\,$}

\def\kms{km s$^{-1}$}

\def\casgm20{CAS-G-M$_{20}\,$}
\def\m20{M$_{20}\,$}

 \def\kms{\,km\,s$^{-1}$}

\begin{document}

\title{Gravitational Waves from Black Holes in Merging Ultra-Dwarf Galaxies}

\correspondingauthor{Christopher J. Conselice}
\email{conselice@nottingham.ac.uk}

\author[0000-0002-0786-7307]{Christopher J. Conselice}
\affiliation{Centre for Astronomy and Particle Theory,
University of Nottingham, University Park, Nottingham, NG7 2RD, UK}

\author{Rachana Bhatawdekar}  
\affiliation{Centre for Astronomy and Particle Theory,
University of Nottingham, University Park, Nottingham, NG7 2RD, UK}

\author{Antonella Palmese}
\affiliation{Fermi National Accelerator Laboratory, P. O. Box 500, Batavia, IL 60510, USA}

\author{William G. Hartley}
\affiliation{Department of Physics \& Astronomy, University College London, Gower Street, London WC1E 6BT, UK}



\begin{abstract}

The origin of the black-hole:black-hole mergers discovered through gravitational waves with for example the LIGO/Virgo collaboration are a mystery. We investigate the idea that some of these black holes originate from the centers of extremely low-mass ultra-dwarf galaxies that have merged together in the distant past at $z>1$.  Extrapolating the central black hole to stellar mass ratio suggests that the black holes in these mergers could arise from galaxies of masses $\sim 10^{5} - 10^{6}$ \solm. We investigate whether these galaxies merge enough, or too much, to be consistent with the observed GW rate of $\sim 9.7-101$ Gpc$^{-3}$ yr$^{-1}$ using the latest LIGO/Virgo results.  We show that in the nearby universe the merger rate and number densities of ultra-dwarf galaxies is too low, by an order or magnitude, to produce these black hole mergers. However, by considering that the merger fraction, merger-time scales, and the number densities of low-mass galaxies all conspire at $z>1-1.5$ to increase the merger rate for these galaxies at higher redshifts we argue that it is possible that some of the observed GW events arise from BHs in the centers of low-mass galaxies. The major uncertainty in this calculation is the dynamical time-scales for black holes in low-mass galaxies.  Our results however suggest a very long BH merger time-scale of 4-7 Gyr, consistent with an extended black hole merger history.  Further simulations are needed to verify this possibility, however our theory can be tested by searching for host galaxies of gravitational wave events.  Results from these searches would will put limits on dwarf galaxy mergers and/or the presence and formation mechanisms of black holes through PopIII stars in the lowest mass galaxies.

\end{abstract}

\keywords{Gravitational Waves, Galaxy Mergers}


\section{Introduction} \label{sec:intro}

The recent discovery of gravitational wave events with LIGO and Virgo have revolutionized many areas of astronomy.  Perhaps the most obvious success of gravitational wave detections to date, beyond the detections themselves,  is the recovery of neutron star mergers through GW170817 in August 2017, which resulted in a slew of studies concerning everything from the formation of elements, cosmology,  the nature of gravity, to neutron star physics, among others (e.g., Abbott et al. 2017; Palmese et al. 2017).  
The event of GW170817 was a landmark one as it involved the identification of a counterpart in electromagnetic radiation that could be followed up.  Amongst the other gravitational wave events to date there has been no confirmed counter-parts or host galaxies, making their discovery interesting mainly due to the fact that gravitational waves were discovered, and the inferred existence of black holes of several tens of solar masses that merged together.  These black-hole/black-hole (BH-BH) mergers remain a mystery, as it is not yet clear how systems as massive as this can form in the first place and then to later merge.  Both the formation and later merging of binary black holes in star formation events ar significant astrophysical problems.  

There are several scenarios in which massive black holes such as these can form and then eventually merge together.  In stellar evolution, stars which are more massive than $\sim 20$ \solm will in principle evolve into black holes, although systems which have masses $>100$ \solm potentially vanish due to pair instability supernova that destroys the entire star (e.g., Fryer et al. 2012).  Furthermore, black holes in X-ray binaries in our Galaxy are known to have masses lower than those of the gravitational wave event black holes (e.g., Casares et al. 2017).  However, theory shows that it is possible to create such binary black holes (e.g., Belczynski et al. 2016; Stevenson et al. 2017) even though none have yet been detected.   

It is also possible that these GW events arise from merging star clusters (e.g., Hong et al. 2018).  Another more speculative idea is that these systems are formed from primordial black holes which are produced early in the history of the universe, which may also account for dark matter (e.g., Hawking 1971).  It remains most likely that many GW events are produced in star formation sites in galaxies given their close proximity, allowing for rapid merging and short dynamical friction time-scales (e.g., Belczynski et al. 2016).  However, other scenarios are worth exploring as not all GW events with black holes necessarily must have the same origin.

The hypothesis we take in this paper is that some black holes seen in the LIGO/Virgo results are the result of mergers of black holes which existed in the centers of necessarily low-mass dwarf galaxies in the distant past.  In more general terms we calculate the likely number of black hole mergers there are at the low-mass end of the galaxy mass function.  To address this question we use a host of astrophysical information, including the galaxy mass:black hole mass relation, the number of lower mass galaxies at different epochs, and the time-scales for how long black hole mergers take to merge once their host galaxies have already merged.  

Furthermore, the implications for gravitational wave events involving black holes at such masses goes far beyond the detection of gravitational waves themselves, but possibly relates to the evolution and formation of galaxies.  One major question is how did these black holes form, and how did they get into a position to merge together?  Were all of these systems formed in star formation regions near each other, or were they formed in separate galaxies?  If the latter is the case for some systems,  this suggests that these events may reveal to us information about the merging and formation of galaxies that is otherwise difficult to impossible to infer from other information. Likewise, ruling this idea out, which in principle is straightforward, will have important implications for central massive black holes in the lowest mass dwarf galaxies.

The LIGO detectors have found 11 gravitational wave events as of late-2018, 10 of which are black hole-black hole mergers with the masses of the merging companions are on the order of 10-70 \solm (Abbott et al. 2018).   This includes LVT151012 which has a 90\% probability of being a real GW event (Abbot et al. 2016).  It is important to ask how, or if, these black hole mergers fit into our picture of galaxy evolution, or can reveal new light on processes which may produce these black holes to begin with. Understanding this will also give us some clues for how to find the host galaxies of these events which otherwise emit, as far as we know, no electromagnetic radiation.    Finding the host galaxies of GW events likely must be deferred until high resolution positioning is available using many gravitational wave detectors to pin-point the location of the host galaxies. It thus might be some time before this idea can be fully tested in the absence of afterglow light.  However, detailed theoretical work, and some observations, can be done to determine whether or not our hypothesis is likely or not.  

The outline of this paper is as follows: \S 2 we discuss the systems we consider in our paper and the data/results we use to analyze their properties, \S3 is an outline of our method for deriving the merger rate of central black holes which may exist in the centers of dwarf galaxies, \S 4 is a discussion of our time-scales and the implications for finding the host galaxy of future systems and \S 5 is our summary.

\section{Data}

The data we use is from a combination of different datasets.    For the gravitational waves we use the information from the latest LIGO/Virgo survey  paper, Abbott et al. (2018). In Table~1 we list the sources considered in this paper, which are the complete set of black-hole/black-hole mergers known from the existing LIGO/Virgo results (Abbott et al 2018).    There is a range of redshifts for these systems from $z \sim 0.1$ up to $z \sim 0.5$ as well as a range of masses for the more massive systems from $m_{1}$ = 10.9 \solm up to 50.6 \solm for GW170729, which is also the most distant of the detected systems.  

If the mass ratios of these black hole mergers reflect the mass ratios of their
host galaxies, then these would also be considered major mergers.  The criteria
for this is that the ratio between the galaxies' stellar masses $M_*$ satisfies $\mu = M_{*,1}/M_{*,2} < 3$. 
 All of our black hole mass ratios are $< 2$, which  makes these systems nearly 
 equal mass major mergers.

Many different data sets are used to derive the properties of the possible merging systems.  
This includes data to determine the likely number densities of galaxies as a function of redshift, as 
well as the merger fraction of these galaxies. We do not have a firm measurement of either of 
these for 
low-mass galaxies at $M_{*} < 10^{6}$ \solm at $z > 0.1$, so they have to be extrapolated 
from estimates at other mass scales.  

For the evolution of the number densities of these galaxies we use the frame-work presented in Conselice et al. (2016) 
who carried out a compilation of all stellar mass functions up to $z \sim 6$ to create a modeled method for deriving the most 
likely mass functions as a function of redshift.    For the merger rates we use the results from Casteels et al. (2014) and Mundy et al. (2017) 
to derive the likely merger history for these ultra-dwarf galaxies. We discuss how this is done in more detail in the relevant 
subsections in \S 3.

    \begin{table}
          \begin{tabular}{c c c c c c}
         
            \hline
            Events & z & m$_{1}$/\solm & $\mu$ & M$_{f}$/\solm & log M$_{*,1}$ \\
            \hline
GW150914 & 0.09$^{+0.03}_{-0.03}$ & 35.6$_{-3.0}^{+4.8}$ & 0.86 & 63.1$^{+3.3}_{-3.0}$  & 5.4$^{+0.53}_{-0.66}$  \\
GW151012 & 0.21$^{+0.09}_{-0.09}$ & 23.3$_{-5.5}^{+14.0}$ & 0.58 & 35.7$^{+9.9}_{-3.8}$ & 5.2$^{+0.55}_{-0.68}$  \\
GW151226 & 0.09$^{+0.04}_{-0.04}$ & 13.7$_{-3.2}^{+8.8}$ & 0.56 & 20.5$^{+6.4}_{-1.5}$ &  5.0$^{+0.57}_{-0.70}$ \\
GW170104 & 0.19$^{+0.07}_{-0.08}$ & 31.0$_{-5.6}^{+7.2}$ & 0.65 & 49.1$^{+5.2}_{-3.9}$ &  5.3$^{+0.54}_{-0.66}$ \\
GW170608 & 0.07$^{+0.02}_{-0.02}$ & 10.9$_{-1.7}^{+5.3}$ & 0.70 & 17.8$^{+3.2}_{-0.7}$ &  4.9$^{+0.57}_{-0.70}$ \\
GW170729 & 0.48$^{+0.19}_{-0.20}$ & 50.6$_{-10.2}^{+16.6}$ & 0.68 & 80.3$^{+14.6}_{-10.2}$ &  5.5$^{+0.51}_{-0.64}$ \\
GW170809 & 0.20$^{+0.05}_{-0.07}$ & 35.2$_{-6.0}^{+8.3}$ & 0.68 & 56.4$^{+5.2}_{-3.7}$ &   5.4$^{+0.53}_{-0.66}$ \\
GW170814 & 0.12$^{+0.03}_{-0.04}$ & 30.7$_{-3.0}^{+5.7}$ & 0.82 & 53.4$^{+3.2}_{-2.4}$ &   5.3$^{+0.54}_{-0.66}$ \\
GW170818 & 0.20$^{+0.07}_{-0.07}$ & 35.5$_{-4.7}^{+7.5}$ & 0.75 &  59.8$^{+4.8}_{-3.8}$ &  5.4$^{+0.53}_{-0.64}$ \\
GW170823 & 0.34$^{+0.13}_{-0.14}$ & 39.6$_{-6.6}^{+10.0}$ & 0.74 & 65.6$^{+9.4}_{-6.6}$ &  5.4$^{+0.53}_{-0.64}$ \\
            \hline
          \end{tabular}
        \caption{The full list of gravitational wave events seen to date with 
the LIGO/Virgo data and their derived properties, including their redshifts, mass of the more massive black hole (m$_{1}$), the ratio of the masses of the black holes merging ($\mu$), the final mass of the black hole after merger (M$_{f}$), and the derived stellar mass of the more massive of the two merging galaxies which potentially produced the black holes (M${*,1}$).}
        \label{tab:stellarhalo}
        \end{table}

\section{Method}

To infer the likely merger rate of central BHs, we need to consider a few observationally based facts. These include: the merger rate of galaxies, the mass of the central black holes in these galaxies, and finally the merger rate of these black holes, or the time-scale of their merging, within the galaxy merger remnant.  All of these quantities are not currently well constrained.  We use a combination of observational results and theoretical modeling to determine what these features are.  In some cases we have no direct measurements of these values, and thus have to make inferences based on the data that we do have, typically using galaxies at higher stellar masses.

We first investigate the black-hole mass/galaxy mass relation, which allows us to answer the question of whether the black holes we see in GW events could possibly arise from black holes which may exist in the centers of low-mass galaxies.  We then investigate the merger rate of galaxies, and the number densities of low-mass galaxies, and thus infer what is the likely merger rate  for the lowest mass galaxies in the nearby and distant universe up to $z \sim 3$.

\subsection{Black Holes in Ultra-Dwarf Galaxies}

The masses of the BHs found by LIGO/Virgo are often several 10s of solar masses.  The question we address in this section is where these BHs are arising from. The most obvious answer is that they arise from black holes which form in star forming regions in galaxies.  The idea here is that massive stars form in star formation episodes undergo stellar evolution and explode as supernova, leaving a core remnant of a black hole.  The question is however what is the likely mass of this remnant black hole?

There are several ways to address this. The first is to empirically examine the masses of black holes in our own Milky Way.  This is certainly incomplete, but what is found is that all of the black holes discovered in the Milky Way, besides the central massive one, contain masses which are $< 20$ \solm.  These objects cannot be the progenitors of the very massive black holes that LIGO/Virgo have discovered (Table~1).  There is in general a lack of evidence that there are intermediate black holes with masses $> 50$ \solm within our own Galaxy.

While there is very little evidence for intermediate black holes, creating these theoretically is also extremely difficult. It is thought that stars form with a maximum mass of around $\sim 100$ \solm.   At least this is a natural limit for producing a remnant when a star evolves off the main-sequence.    If stars were more massive than this, they would explode in pair-instability supernova, leaving no black hole or any remnant.   However, when these $<100$ \solm stars supernova they do not retain all or even a significant amount of their mass which leaves a small fraction left to form a black hole (e.g., Limongi \& Chieffi 2018). 

In fact, in some metal rich cases it is impossible to reach the mass limit of the LIGO black hole detections from stars that have gone supernova.  Whilst
metal poor conditions can produce higher mass systems, these are rarely seen in these nearby universe for massive galaxies.  Only a few low rotational models are able to predict black holes remnants with the masses detected by LIGO (e.g., Limongi \& Chieffi 2018).

It is also the case that for a merger to occur between two black holes, they need to be quite close to each other when they are born.  However, the giant phase of stellar evolution means that those stars close to enough to merge would become a single systems before they were black holes.  To relieve this requires that the binary stars which become binary black holes would have to have separations implying a merger time-scale of 100 million Hubble times (e.g., Celoria et al. 2018).  Furthermore, a supernova explosion, necessary to create black hole remnants would disrupt these binary systems. 

This leads us to consider how the typically metal poor environment of low mass galaxies may lead to the formation of stellar mass black holes.  We first investigate this by examining the likelihood of there being massive black hole formed in star formation events followed up supernova.  The question here is whether there would be stars within dwarf galaxies that would become massive black holes with masses $> 30$ \solm.   If we assume that only massive stars with a Salpter Initial Mass Function will survive to become black holes with masses that LIGO has identified, we are left with up to a hundred or so stellar-mass sized black holes within galaxies of mass M$_{*} \sim 10^{6}$.  These black holes if they exit could be one route to form mergers as seen in LIGO events.  Dwarfs are a likely place for this given that they have lower metallicities as opposed to higher mass galaxies which are often or even always more metal rich.  


We then consider  the black hole galaxy mass relation 
and whether there is a consistency that the black hole masses measured in 
GW events could arise as merging central black holes in ultra-dwarf galaxies. 
Whilst it is true 
that central black hole masses are often quite high, around $10^{6}$ \solm 
for the highest mass galaxies, recent results suggest that central black 
hole masses for dwarf or low-mass galaxies are {\em less massive} than what 
would be inferred 
from the high end of the black hole mass:galaxy mass relation (BH:GM) (e.g., Reines \& Volonteri 2015; hereafter RV15).  It
is also becoming clear that dwarf galaxies contain central black holes and
AGN activity (e.g., Reines et al. 2013; Baldassare et al. 2015).  

The extent of this is not entirely clear yet at even lower masses as these systems have hardly been
studied, thus inferences have to be made until better data arrives.   Furthermore, as shown in RV15 and other papers, the scatter of the BH:GM relation becomes larger at lower masses, implying that some low-mass systems must have black holes in the range of $<100$ \solm.

The quantitative relation of BH mass to galaxy mass is well calibrated for higher mass galaxies, but for lower mass systems it is still not well defined.   By examining low-mass dwarf systems RV15 find a black hole galaxy mass relation such that that,

\begin{equation}
{\rm log} (M_{\rm BH}) = \alpha + \beta {\rm log} (M_{*}/10^{11} {\rm M_{\odot}\,}),
\end{equation}

\noindent in solar units for the masses M$_{*}$ and M$_{\rm BH}$.  The constants fit by RV15 are $\alpha = 7.45\pm0.08$ and $\beta = 1.05\pm0.11$.   This relationship allows us to relate the black hole masses measured from the LIGO/Virgo results to the inferred masses of these galaxies which potentially hosted these BHs.   Note that we do this as simply a test to see if our hypothesis has any validly whatsoever. It is unlikely that the RV15 relation holds exactly at such low masses, however the trend is such that the relation would only likely get steeper, meaning that it would be in principle possible to find low-mass BHs in ultra-dwarf galaxies.

Thus, using this relation if the black holes we detect from the LIGO/Virgo results are due to black holes in the centers of galaxies involved in
galaxy mergers we can calculate their original host galaxy mass.    
The masses of the BHs merging in GW events  range from 10.9 \solm to 50.6 \solm (e.g,. Mandel \& Farmer 2018; Abbott et al. 2018; Table~1).  We show on Figure~1 the inferred relation between the derived stellar mass of a galaxy and the central black hole mass, extrapolating to lower masses via the RV15 relation in Table~1.  The horizontal lines show the masses of the more massive of the LIGO/Virgo  black holes for each merger, and the connecting vertical line shows the stellar mass of the host galaxy derived using the relation above.  The uncertainties for this relation are shown by the dashed blue lines.   
This gives us a range of possible masses of the host galaxies for these sources of $M_{*} = 10^{4.5} - 10^{6}$ \solm. While these are low-mass galaxies, they are not unheard of, and in fact probably dominate the universe in terms of numbers (e.g., Conselice et al. 2016).  Galaxies with these masses, or lower, are also expected to form in the 
universe at  $z > 10$ (e.g., Tegmark et al. 1997) and are seen in the Local Group.  In the following, we use these results to determine the co-moving volume number densities of galaxies within this mass range, which in turn is a necessary ingredient to infer the galaxy merger rate. 

        \begin{figure}
          \centering
          \includegraphics[angle=0, width=8.5cm]{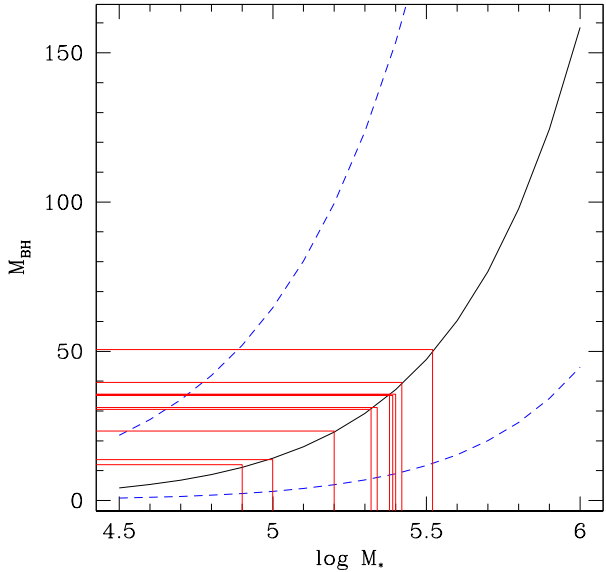}
          \caption{The relation between black hole mass and galaxy mass
as taken from Reines \& Volonteri (2015), extrapolated to lower masses than 
those used in that study.  In general Reines \& Volonteri (2015) find that 
central black holes masses are lower for their given host galaxy mass,
compared to high mass galaxies, at the lower mass end of
their studied relation, finding black holes masses down to 10,000 \solm.  
We show using the red lines the black-hole masses of the more massive member 
 measured in the
LIGO/Virgo detections and the derived mass of the host galaxy for each.
The solid black line is the primary relation derived, and the blue dashed
lines show the 1 $\sigma$ error on this best fit. }
          \label{fig:abundance}
        \end{figure}

\subsection{Merger Rates of Low-Mass Galaxies}

To investigate the potential merger rate of black-holes in lower mass galaxies, we consider the merger rate of their potential host galaxies.  As mentioned earlier if these massive black holes were formed in the central portions of galaxies, then based on the galaxy mass-black hole mass relation given in eq. (1) we would expect the hosts of these lower mass central black holes to be low-mass galaxies.  We next derive the merger rate for these systems, as this will ultimately be a major clue towards understanding if mergers of lower mass galaxies can produce some of the GW events  found to date and potentially future events with other detectors.  Overall, the galaxy merger rate (number of merging events per Gyr per co-moving Mpc$^{3}$; Conselice 2014) can be written as:

\begin{equation}
\Gamma_{\rm GM} (z) = \frac{f(z)}{\tau(z)}{\phi(z)},
\end{equation}

\noindent where we have a mixture of observationally and theoretically derived quantities. The function $f(z)$ is the redshift--dependent galaxy merger fraction, while $\phi(z)$ is the evolution of the volume number density.  Finally, the time-scale for merging, $\tau (z)$, is a quantity that must be derived from theory (e.g., Mundy et al. 2017).  Note that our merger fraction is defined to be the number of mergers per galaxy (not the fraction of galaxies merging, which is different by a factor of $\sim 2$).  This is done to mimic the LIGO/Virgo rates which are merger events, not the number of black holes merging.  

First, we examine the likely merger fraction of these galaxies, as this will reveal whether there are enough to account for the GW events.    The merger fraction and rates for galaxies at such low masses have not been measured. The galaxy merger fraction and rates for the lowest mass galaxies are taken from a number of sources.  The lowest mass galaxies for which the merger rate has been measured are taken from Casteels et al. (2014), however we also consider the results of low mass galaxy mergers from the GAMA survey as described in Mundy et al. (2017).  

The nearby galaxy merger rate at the lowest masses ($\sim10^{8}$ \solm) is measured as $\sim 0.02$ mergers Gyr$^{-1}$.    This is based on merger time-scales from Lotz et al. (2010) who determine merger rates based on numerical N-body models which include star formation and feedback physics (Casteels et al. 2014).  
We will thus make the assumption that the merger rate for lower mass galaxies is at the same level. Galaxies at these low masses cluster at the same scale, or possibly even more strongly, for example they are often found in rich clusters of galaxies (e.g., Penny et al. 2015).  Therefore it would appear that
this assumption is likely valid.  Our assumption also follows in general what is predicted in theory for the merger rate of low-mass galaxies (e.g., Snyder et al. 2017).

Furthermore, the merger fraction $f(z)$ increases with look-back time. This is necessary to consider as the time-scales for BH mergers, after their host galaxies has merged, can be several Gyr long (e.g., Mapelli et al. 2019), thus this is not a process which starts and completes within the local universe.  The typical way in which this is represented is through a power-law increase of $(1+z)^m$, thus we represent the galaxy merger fraction evolution as:

\begin{equation}
f(z) = f_{0} \times (1+z)^{m}
\end{equation}

\noindent where $m$ is the power-law index and $f_{0}$ is the local or $z = 0$ merger fraction for our low-mass galaxies.

We also need to understand how the time-scale for mergers, or the merger rate for galaxies, changes at higher redshifts.  This is because if the time-scales for mergers were faster in the past then there would have been many more mergers, at a given merger fraction, compared with lower redshifts. As shown by Snyder et al. (2017) the time-scales for mergers decline as $\tau (z) \sim (1+z)^{-2}$ when probing higher redshifts.    We thus implement this evolution in the merger time-scale which does a good job of matching the observed and predicted merger rates at high redshifts (e.g., Duncan et al. 2019).

Next, we need to consider the evolution of the number densities, $\phi(z)$, of these lower mass galaxies, which can potentially host the black holes producing the LIGO/Virgo GW events from  black hole mergers.  We know that the slope $\alpha$ of the power-law or Schechter function fit of the mass function becomes steeper as we go to higher redshifts (e.g., Duncan et al. 2014; Conselice et al. 2016; Bhatawdekar et al. 2019).  

The number density evolution, as discussed later in \S 3.4 can thus be represented by a power-law fit of the form:

\begin{equation}
\phi(z) = \phi_{0} \times (1+z)^{q}
\end{equation}

\noindent where $q$ is the power-law index for the increase in the number densities of lower mass galaxies which we probe at higher redshifts.  
Putting this all together we find that the galaxy merger rate $\Gamma$ can be represented by:

\begin{equation}
\Gamma_{\rm GM} (z)\,  = \frac{f_{0} \phi_{0}}{\tau_{0}} (1+z)^{(2+m+q)}.
\end{equation}

\noindent  There are thus five unknowns in this equation which can be derived or inferred based on observational estimates.

\subsection{Parameterizing the Galaxy Merger History (m)}

The merger history for low-mass galaxies, such as the ones that may produce the GW BH mergers, is unknown.    However, what has been shown is that for all galaxy types there is an increase in the merger fraction such that the exponential on the power-law varies between $m = 2-3$ (e.g., Mundy et al. 2017).  We thus use the best-fitting values for the increase in the pair and merger fraction from Mundy et al. (2017), using a value of $m = 2.68^{+0.59}_{-0.59}$ from the best fit for all samples at high redshift up to $z \sim 3$.  See Mundy et al. (2017) for details on how this is calculated and computed using data from the three deepest extragalactic near-infrared fields.  

When we examine the merger history in simulations, such as from  the Illustris simulation (e.g., Rodriguez-Gomez et al. (2015), we finds similar results to the data.  We therefore use this exponent, and its error ranges to determine the number of mergers our low-mass galaxies  undergo which potentially host the black holes which merge to produce the GW events.

\subsection{Number Density Evolution (q)}

We also have to extrapolate the number densities of these low-mass galaxies and how they evolve at higher redshifts.  This can be done through the same formalism that we used to determine the total number densities of all galaxies at high redshifts in Conselice et al. (2016). When integrating the number densities between log ($M_{*}/$\solm )$ = 4.5 - 6$, we get the result shown in Figure~2 using
data from various surveys at all redshifts.  To fit the evolution of $\phi(z)$ we carry out a MCMC analysis using a pure-Python implementation of Goodman \& Weare's Affine Invariant Markov Chain Monte Carlo (MCMC) ensemble sampler (Goodman \& Weare 2010; 
Foreman-Mackey et al. 2013).  We carry out our analysis to determine the 
evolution of the number density with redshift using eq. (4), and fit for the parameters in that
equation.

        \begin{figure}
          \centering
          \includegraphics[angle=0, width=8.5cm]{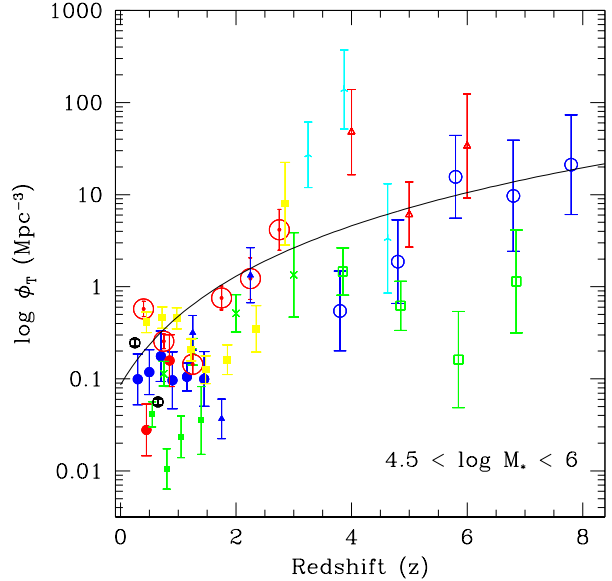}
          \caption{The number densities of galaxies with stellar masses
between log M$_{*} = 4.5$ and log M$_{*} = 6$.  These are taken from the
mass function computation by Conselice et al. (2016).   The solid line is
the fit to this relation as discussed in \S 3.4. For the most part,
particularly at higher redshifts, this is an extrapolation from the lowest
limits in which these mass functions are directly measured (see Conselice
et al. 2016) for details and for the symbol type definition.}
          \label{fig:abundance}
        \end{figure}

        \begin{figure}
          \centering
\vspace{-2cm}
            \includegraphics[angle=0, width=8.5cm]{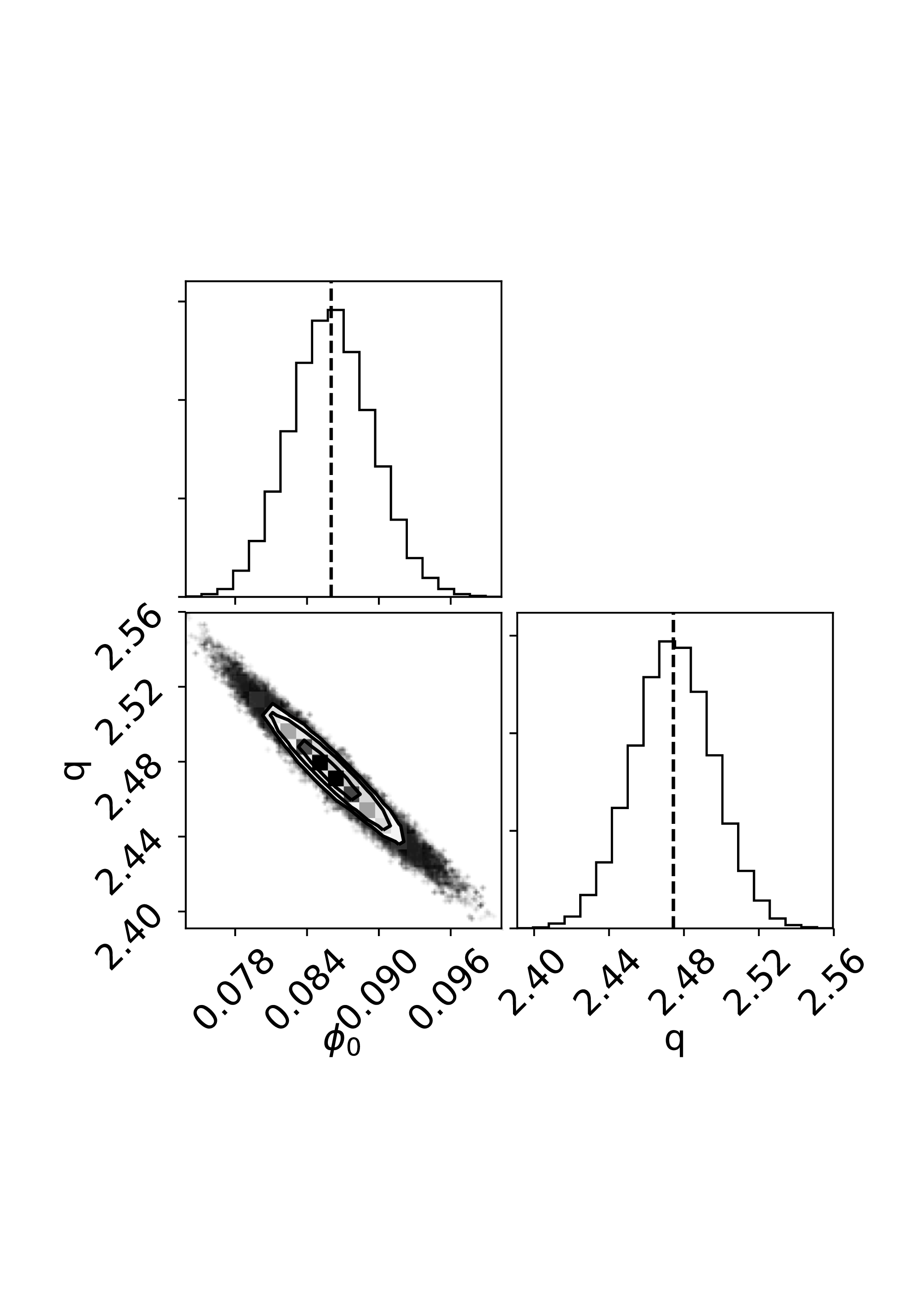}
\vspace{-2cm}
          \caption{Plot of the contours and distributions of 
posteriors for the MCMC fit of the evolution in number densities using
the data in Figure~2.  As can be seen we find a fairly stable fit for
both the central density for low-mass galaxies, $\phi_{0}$ and the power-law
exponent for the increase in density, $q$.  }
          \label{fig:abundance}
        \end{figure}

\noindent We have inferred the best fit values for $q$ and $\phi_{0}$ using 10$^{2}$ MCMC chains of 10$^{4}$ steps each.  
The posterior values for this fit are showed in Figure~3 and has a slope of $q = 2.47\pm 0.02$, and $\phi_{0} = 0.086 \pm 0.003$.  This
shows that the number density, $\phi$ of these low-mass galaxies increases significantly as we go to higher redshifts.  It also implies 
that there were more low-mass galaxies at early times which must have merged together to form the ultimate black hole mergers that could produce
the GW events.

\section{Calculation Result}

\subsection{The Evolution of Low-Mass Galaxy Mergers}

The final result of our calculation of the merger rates for low-mass 
galaxies, plotted
in terms of events per Gpc$^{3}$ per year, is shown in Figure~4.  There are 
two ways in which we show this relation between the number of merger events and 
redshift.  The first is the solid line which shows the number of expected 
merger events if we use the local values of the merger history of local 
galaxies, and make the assumption that the merger-time scale declines at 
higher redshifts, and the number densities of galaxies at this mass
range remains the same.  The dashed line shows the relation between the
number of merger events and redshift when we evolve the number densities 
using the relations discussed in \S 3.4.

We also show the range of possible gravitational wave event rates by the dashed
blue horizontal lines going from a few events per year per Gpc$^{3}$ to almost
1000 (e.g., Abbott et al. 2016).   More recently this rate has been updated
to 9.7-101 Gpc$^{-3}$ yr$^{-1}$ (Abbott et al. 2018) which we show as the
solid red lines.

        \begin{figure}
          \centering
          \includegraphics[angle=0, width=8.5cm]{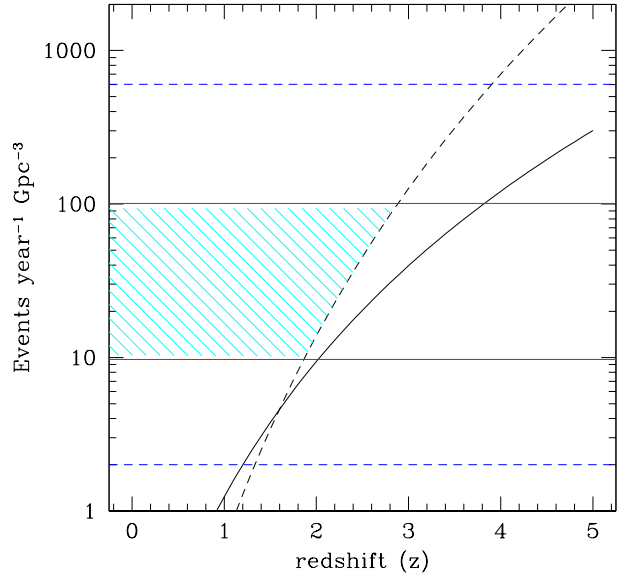}
          \caption{The rate of merging events is shown as the two lines (dashed
and solid) as a function of redshift in units of events per Gpc$^{3}$ per 
Gyr.  the blue dashed horizontal lines show the limits of the gravitational wave 
events from Abbott et al. (2016), while the red solid lines show
the updated constraints from Abbott et al. (2017).  The hashed area gives the possible
range where the merger events are similar to the GW events.    The
solid line shows the evolution of the merger rate if we assume merger
time-scales decline with redshift, but that the number densities of
galaxies at high redshift is similar as it is today.  The dashed line
shows this same relation when we derive the likely number densities of 
these galaxies using the relations described in \S 3.4.}
          \label{fig:abundance}
        \end{figure}

 This figure shows that we are able to
match the gravitational wave event rate at about $z \sim 1.5$ for the best-case scenario.  If
the higher limit is used, then the merger rate only matches 
at $z \sim 2.5$.  This higher merger rate is largely due to the increase
in number densities of these lower mass galaxies at higher redshifts, as
well as the decline in the  merger time-scale.  It now remains to
be seen, and to show, that it is possible for the delay between the
merger of two galaxies and the later merger of their central black holes is 
similar to the time between the galaxy merger event at $z \sim 1.5$, or higher, and when the GW
is observed.  Since these GW events are between $\sim 300-2200$ Mpc in luminosity distance, this means that the 
GW wave was produced between 1 to 5 billion years ago.  Below we carry out a likelihood calculation 
for what is the time-scale for the merging of the central black holes in these merging systems.

\subsection{The BH-BH Merger Time-Scale}

\subsubsection{BH Merger Time-Scale Calculation}

The GW events we see from LIGO/Virgo occurred a few Gyr ago.  The question is: when were the black 
holes that produced these events created and how long did it take for them to merge?   

There are not many calculations or simulations of the  likely time it takes for two central black holes to merge in low-mass 
galaxies, yet there are some ideas and calculations we can use to create a simple model. There are two scenarios which we investigate in this paper.  The first is the relaxation time for a black hole within a new system such as a merged ultra-dwarf galaxy by interacting with similar mass black holes that may exist from past supernova. The second is the dynamical friction between the existing black holes and stars within the ultra-dwarf galaxies and the new black central black hole which has entered.  

Through various arguments from e.g., Binney \& Tremaine (1987) it can be shown that the tw-body relaxation time-scale for a system with mass $M$, size $R$, and $N$ number of particles goes as:

\begin{equation}
t_{\rm relax} = 10^{8} {\rm yr} \left(\frac{1}{{\rm log} N} \right) \left(\frac{M}{10^{5} M_{\odot}\,} \right) \left(\frac{R}{\rm 1 pc} \right)^{3/2} \left(\frac{1 M_{\odot}\,}{m}\right)
\end{equation}

\noindent where $m$ is the mass of the black hole of interest (e.g., Celoria et al. 2018), in our case $\sim 50-70$ \solm. Ultra-dwarf galaxies have been discovered in the Local Group such as Segue 2, Ursa Major II, Leo IV and Leo V.  These galaxies have low metallicities [Fe/H] $\sim -2.5$, absolute magnitude and masses similar to the level at which we are examining mergers in this paper, with absolute magnitudes M$\sim -5$ and brighter, with velocity dispersions from 30 \kms to 100 \kms, and sizes which are $\sim 100$ pc (e.g., Walker et al. 2009).  

 Using the values for known ultra-dwarf galaxies, we calculate that the relaxation time-scale is on the order of a few Gyr, certainly less than a Hubble time.    This is opposed to large mass galaxies such as those expected to host supermassive black holes whereby the relaxation time is up to 10,000 times the Hubble-time.  Ultra-dwarf galaxies thus provide a unique environment for $\sim 50$ \solm black holes to become dynamical relaxed by the $\sim 100$ of similar mass black holes already present within the ultra-dwarf.

Second, simple dynamical friction time-scales in shallow profiles suggest that the time for two black holes in a dwarf galaxy to merge
is between 10-100 Gyr  (e.g., Binney \& Tremaine 1987), depending on orbital eccentricity,  this is
certainly too long to be a viable path for GW events.  We investigate this in more detail below.  
The dynamical friction time-scale can be given by:

\begin{equation}
t_{f} = \frac{1.17}{\rm ln \Lambda} {\rm Gyr} \left(\frac{r_{i}}{\rm 5 kpc} \right)^{2} \left(\frac{\sigma}{\rm 200 km s^{-1}}\right) \left( \frac{\rm 10^{8} M_{\odot}\,}{M} \right)
\end{equation}

\noindent where $r_{i}$ is the initial distance, $\sigma$ is the velocity dispersion of the galaxy of interest and $M$ is the mass of the black hole.   Using typical values for ultra-dwarf galaxies in the nearby universe we use values of $r_{i} \sim 100$ pc, $\sigma \sim 300-100$ \kms.  By using a value of $M \sim 50$ \solm for the largest LIGO black hole, we indeed find a time for dynamical friction of $t_{f} \sim 100$ Gyr, obviously far too long to produce a merger.

 A simple application of dynamical friction is however not necessarily applicable in our situation as
dwarf and low-mass systems can, and do, differ from more massive galaxies.      The lowest mass dwarfs are very dark matter dominated and very small so simple scaling may not directly apply. 
Simulations do show that mergers for low-mass dwarf galaxies 
take longer to occur than higher mass black holes.   However, this time-scale increase does not scale as steeply as simple dynamical friction calculations would suggest, and thus it
seems possible that lower mass dwarfs would also have time-scales on the order of a Hubble 
time or less.

There is also the fact that many of these central black holes will likely be embedded in central star clusters (e.g., Seth et al. 2010).  These star clusters
are such that their extra mass would provide a possible conduit to allow faster dynamical friction to occur, thereby leading to
more rapid merging of low-mass black holes in the centers of ultra-low mass dwarfs.    This envelope and extra mass would guide the black hole and protect it until it
reached the center of the merging system whereby it will merge with the 2nd black hole.  In fact recent simulations suggest that this is not only a sufficient, but a 
necessary method for producing black hole mergers (e.g., Antonini \& Rasio et al. 2016; Pfister et al. 2019).  

In fact, using the above equation with a mass of 1000 \solm we obtain a dynamical friction time-scale of $< 10$ Gyr.  Thus a star cluster surrounding a black hole in these systems which is only a factor of 10 more massive would be sufficient to allow the black hole within the star clusters to reach the bottom of the potential well of the remnant merger.    In fact, because these black holes are low mass, they will more readily reach a smaller separation at sub-pc scales before they begin to 'see' each other during a merger, thus facilitating a rapid sub-pc merger.    This requires less hardening with regards to the stellar background, and thus naturally leads to a decline and merger at sub-pc scales due to gravitational wave radiation.  

Furthermore, black hole mergers in dwarf galaxy simulations, for systems several orders of magnitude larger than what we consider, show that the time-scale for these mergers is on the order of $\sim 6-8$ Gyr (Tamfal et al. 2018), similar to what we need in our scenario.  Furthermore, some calculations show that the epoch when the black holes that produce GW events occurred should be at high redshift if they originate in star formation events (e.g., Emami \& Loeb 2018).   As Tamfal et al. (2018) show, the profile of the dark matter is critical for determining the time-scale of merging BHs in merging dwarf galaxies.  The general dark matter halo profile can be described by an exponent on the density given by $\gamma$ (e.g., Lokas 2002; Tamfal et al. 2018).  Tamfal et al. (2018) investigate the
time-scales for mergers when $\gamma = 1, 0.6$ and 0.2.  $\gamma = 1$ is an NFW(Navarro, Frenk \& White 1997) profile, while $\gamma < 1$ is more commonly seen in lower mass galaxies (e.g., Oh et al. 2015).  Note that Tamfal et al. (2018) consider intermediate mass black holes (IMBH; $\sim 10^5$ \solm), as central black holes of merging dwarf galaxies that are typically more massive than what is considered in our analysis. As such, we only consider their results a rough time-scale for our case, which does not currently have appropriate simulations available. Future work simulating merging black holes in the lowest mass galaxies would be very revealing and help shed light on this issue.

\subsubsection{Derived BH Merger Rates}

As we have discussed, simulations of merging dwarf galaxies with black holes find that the merger time-scale is long due to less effective dynamical friction.  This time for merging is $> 7$ Gyr in this simulation.   It is also well known that BH mergers often stall at a few pc, and this can lead to quite long merger times of up to 10 Gyr even within dense and massive galaxies (e.g., Volonteri et al. 2015).   As Tamfal et al. (2018) further show, if the dark matter profiles of these galaxies are steep then this leads to a more effective merger and shorter time-scale.   If these simulations can be extended down to even lower masses, then it is entirely possible that merging dwarf galaxies can produce the gravitational wave events seen with LIGO/Virgo in principle, especially if embedded star clusters are considered.

 
        \begin{figure*}
          \centering
          \hspace{0cm} \includegraphics[angle=-90, width=14.5cm]{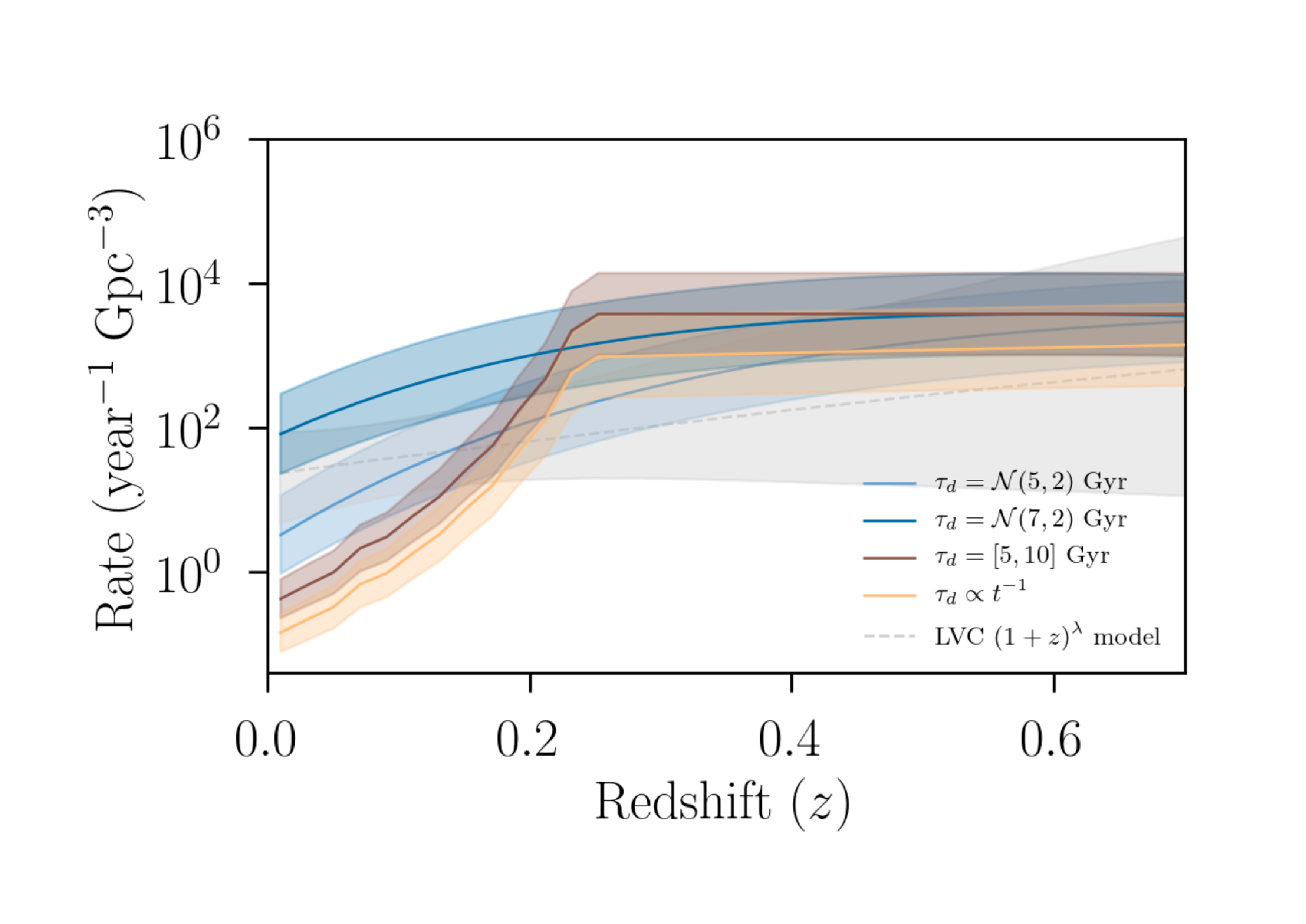}
\vspace{-1cm}
          \caption{Rate of binary black hole mergers produced by dwarf galaxy mergers as a function of redshift $z$, for different time delay distributions.  Here we have used a Gaussian distribution around 5 and 7 Gyr with a standard deviation of 2 Gyr (Cyan and Blue, respectively), a top-hat function between 5 and 10 Gyr (purple) and a distribution that declines as $t^{-1}$ (orange).  We also show a LVC model from Abbott et al. (2018) as the dashed line.  }
          \label{fig:abundance}
        \end{figure*}

We use the delay time-scales discussed as a guide to infer the rate of binary-black-hole mergers at cosmic time $t$ produced by galaxy mergers given some time delay distribution $p(\tau)$ between the galaxy merger and the black-hole merger. The black-hole merger rate $\Gamma_{\rm BHM}$ will follow the relation:

\begin{equation}
    \Gamma_{\rm BHM}(t) = \int_0^t
    {\rm d}\tau~\Gamma_{\rm GM}(t-\tau)~p(\tau)\, ,\label{eq:BHMrate}
\end{equation}

\noindent where $\Gamma_{\rm GM}$ corresponds to the galaxy merger rate in Eq. 5. The BH merger rate evolution with redshift will depend on the model chosen for the time delay. Antonini et al. (2015) and Tamfal et al. (2018) show that the typical time-scales for merging of central black holes, when negligible amounts of gas is present in the merging galaxies, is $\sim 5$ and 7 Gyr, respectively. We test different time delay distributions: a Gaussian $\mathcal{N}(\tau;\mu_\tau, \sigma_\tau)$ around these values with a standard deviation of 2 Gyr, a top--hat function between $[5,10]$ Gyr and a $t^{-1}$ distribution between $[1,10]$ Gyr. Figure~5 shows the rates for these distributions, which can be compared to Figure~4. 

While sophisticated simulations tailored for the particular galaxy and black hole mass range studied here are needed to identify more realistic time delay distributions, it is clear from Figure~5 that this BH merger scenario predicts a rate of BH mergers that evolves with redshift. In particular, the rate is expected to rise with redshift, as a result of the increasing galaxy merger rate with redshift.  Such a trend is currently consistent with what is found for the first two LIGO/Virgo observing seasons (Abbott et al 2018b) (their Figure 5).

\section{Discussion}

We have shown that it is possible in principle that merging ultra low-mass dwarf galaxies
have merger rates which are high enough to produce the gravitational wave events seen by 
LIGO/Virgo.  This assumes that either black holes that exist in these galaxies from star 
formation events or from their being central black holes then merge together in the remnant galaxy.
 However, to match the GW event rate requires that the galaxy merging events occur at
$z > 1$.  Since the GW events are observed at lower redshift, this implies a 
time-delay between these galaxy mergers and the BH mergers of between 6-8 Gyr. This time-scale and
whether BH mergers in dwarfs can occur within this time is likely the biggest uncertainty in 
making this a viable channel for GW production.  There are a couple of implications for 
these results which we briefly discuss here.

 First, as described earlier in the paper, the BH merger ratio is quite close to 1:1, and never 
less than 1:2.  Major galaxy mergers which we compare rates with have merger ratios which range 
from 1:4 to 1:1.  It is likely that the reason we are seeing only similar mass BH mergers is 
that these produce the types of signals that LIGO/Virgo are most likely able to detect.  This however
implies that the merger rate for these systems would be higher if we could detect more `minor' 
BH mergers.  However, for galaxies, most of the mergers are also between galaxies of similar 
mass, not quite close to 1:1, but within a factor of two.  This suggests that the LIGO/Virgo 
events cannot all be produced in BH mergers from low-mass galaxies or else
the nearly 1:1 ratio for galaxies would be slightly lower. However, this does not discount 
that a significant fraction of GW events could arise from these types
of mergers, but it is unlikely to account for all of them.

\subsection{Implications for Host Galaxies of GW Events}

The scenario presented here also has implications for finding hosts of the GW events.  One of the major goals of GW 
studies is to find the host galaxies in which these BH mergers occur, and ultimately in what regions within galaxies
these events arise.    Our results suggest that in the 
absence of a electromagnetic counterpart (e.g., Loeb et al. 2016) the best way to find the sources 
of these mergers is through the properties of their inferred host galaxies if they form from mergers 
of central BHs in low-mass galaxies.  As such, searches for host galaxies even at modest
redshifts will have to probe very deep to find these faint dwarf systems and it is possible that
any `afterglow' would appear as an `orphan' without any obvious host galaxy even in deep imaging. 
This is due to the resulting host galaxy being  low-mass 
systems with masses $\sim 10^{5-6}$ \solm.    However, these host galaxies
would likely retain no galaxy-galaxy merger signatures in their central parts given the long multi-Gyr time-scale which is ample 
enough for the merged host galaxy to morphologically relax.  

An issue with this however that the host galaxy
magnitude will be extremely faint with apparent magnitudes of B $\sim 31-33$ for M$_{*} \sim  10^{6}$ \solm galaxies at $z \sim 0.3$, 
given a reasonable mass to light ratio for these galaxies.   This is a lower mass
and fainter limit than we can probe even with the deepest Hubble Space Telescope
imaging (e.g., Conselice et al. 2011).  Thus, in this scenario only the nearest GW source host galaxies would have a realistic chance to be seen before JWST.  
The large area in which sources can be localized within using LIGO, at least of the order of 10 deg$^{2}$, presents 
another problem as finding these faint sources are like finding a microscopic needle in a cosmic haystack.  This might however
be easier if these sources are being gravitationally lensed, magnifying the
host galaxy of the gravitational source (e.g., Smith et al. 2018).

Since dwarfs are often found
in dense environments, a good possible location for GW source positions would be in clusters of galaxies or groups with satellite galaxies merging.  Mergers between low-mass galaxies would not occur frequently within extreme or low density environments.  Imaging such as with the Dark Energy Survey (e.g., Doctor et al. 2018) would have trouble identifying a host galaxy within this scenario, and finding a counterpart with no host galaxy would give some evidence for this idea.   Unfortunately, LSST at a nominal depth of magnitude
27 will also not be able to identify these hosts.  The only real possibility will be through deep JWST imaging unless these systems are very nearby.

Furthermore, finding black holes in ultra-dwarf galaxies would be another way to determine whether
of not or theory is viable.  This could be done in a number of ways, including deep kinematic observations and perhaps X-ray and radio techniques to try and identify active black holes that might be accreting matter in these galaxies.  However, since few of these galaxies have gas in them, this might prove difficult. Perhaps the best way to test this idea is to search for star clusters within the centers, or near the centers, of these ultra-dwarf galaxies which would have masses $> 1000$ \solm.  This could in principle be carried out today with existing facilities.  Furthermore, finding and studying these ultra-dwarf galaxies at higher redshifts, and in environments other than the Local Group, is another way to make progress in testing this theory.

These GW events would also be found in or near the centers of these galaxies if follow up imaging was deep enough to find these host sources.  If GW events are located in massive galaxies, especially outside their centers and within star forming regions, it would be a difficult observation to reconcile with the idea that these objects formed in low-mass galaxies that later merged.    However if this is the case then some assumptions about dark matter, black hole stellar mass relations, or merging within dwarf galaxies need to be revised.  

If it turns out that no sources of GW waves are within merging or the remnants of merging low-mass
galaxies
then this would imply one of three things: (1) Ultra low-mass dwarfs do not contain central black holes, 
(2) the merger-time scale for BHs in ultra low-mass dwarfs are
longer than what we find in simulations for more massive dwarfs, or (3) that the merger rate of dwarf
galaxies is not as high as we think. All of these have 
implications for our
understanding of dark matter and galaxy/BH  evolution/formation. 
If some GW events are formed in mergers of dwarfs it would be a strong indication that initial black hole formation occurs from seed Population III stars, as opposed to
a collapse of gas early in the universe (e.g., van Wassenhove et al. 2010).

Another way to distinguish this scenario from the star formation one is to look at the spin of the black holes that produce these GW events.  Binary formation scenarios, such as in star formation episodes, predict that the spins of the merging BHs should be more or less aligned (e.g., Piran \& Hotokezaka 2018).  However, what is found is that there is an isotropic distribution in spins for the known BH mergers (Piran \& Hotokezaka 2018), which would be expected for mergers of black holes originating from the centers of different galaxies.   Future observations of black hole mergers, especially in the next few years will help clarify many of these issues.

\section{Summary}

We investigate the possible progenitors and formation mechanisms of the merging black holes seen in the LIGO/Virgo GW events known to date, and possibly future ones with new detectors.   We know that the black holes discovered to date are fairly massive, several 10s of solar masses in mass,  and were previously in a position to merge.  We discuss in this paper the possibility that these massive black holes arise from the mergers of BHs which were in low-mass merging dwarf galaxies.  Since no host galaxy for these events has been found to date, due in no small part to it being very difficult to impossible to identify the source of the GW events based on the emission of an afterglow or counterpart, it remains a mystery where these sources arise from.  However even if all LIGO/Virgo sources are in star forming regions it is still possible that some future events can occur
through alternative channels.   LIGO/Virgo will remain an 
important source for finding possible low-mass galaxy mergers, as opposed to higher mass galaxy mergers better detected with LISA and 
Pulsar Timing Arrays (e.g., Chen et al. 2018).  

Since the source of these events is debated, it is important to address the question of whether these GW sources could arise from binary black holes that were once at the centers of two distinct low-mass ultra-dwarf galaxies that later merged. We addressed this question in this paper by investigating the volume number densities and likely merger history of low-mass galaxies at high redshifts.  We find that the merger rate of galaxies goes up significantly with redshift, as does the number densities of low-mass galaxies.  In summary we find:

\noindent I. Extrapolating the black hole mass-galaxy mass relation calibrated for dwarf galaxies to the lower limits of possible galaxies, we determine that it is possible that low-mass galaxies between M$_{*} = 10^{4.5} - 10^{6}$ \solm could host tens of solar mass black holes in their centers.  This is especially the case if the scatter in this relation becomes larger at lower masses, ensuring that a large fraction of galaxies with these masses host central black holes with masses $< 100$ \solm.  

\noindent II. The merger rate of nearby low-mass galaxies is certainly too low to produce the GW event rate from LIGO/Virgo.  Therefore, if these events are produced by the mergers of black holes in low mass galaxies, the mergers of the galaxies themselves must have occurred in the distant past.

\noindent III. We determine the merger rate and number densities of the lowest mass galaxies in the distant universe.  We combine these results to determine the merger rate for low-mass galaxies up to $z \sim 3$.  We determine that it is possible to reach the lower limits of the GW event rate by $z \sim 1.5$.  This however requires the merger time-scale for the black holes within merging galaxies to be $\sim 6-8$ Gyr.   This time-scale is indeed found in detailed N-body models of BH mergers in merging dwarf galaxies, although the systems studied to date in simulations are all several orders of magnitude larger than the ones we consider here.  However, if these BHs are embedded in massive central star clusters, this would be an effective conduit to drive these systems to the center of the merger remnant where they can merge within $< 10$ Gyr based on dynamical friction and two-body relaxation arguments.  

This time-scale is similar to what found within intermediate mass black-holes in slightly more massive galaxies, but detailed simulations tailored for the galaxy and BH mass ranges studies here will be required to confirm whether this is a reasonable time delay. If this is the case, we expect the rate of BH mergers to increase with redshift for a number of delay time distributions. 

Future observations of GW rates will either give credence to this idea or rule it out. In either case, there will be interesting implications for galaxy formation and evolution, and possibly dark matter. 
This paper also suggests that `orphan' afterglows of GW events should be searched for, as the hosts of these events will be fainter than B$\sim$ 30.   If this scenario is correct then deeper imaging with JWST may be needed to reveal these galaxies for those even at modest redshifts.

In the future, as the resolution and ability to pinpoint locations of GW events improves we will one day be able to determine the location of the  host galaxies of these events.  This will shed light on this problem and either give credence to this idea or rule it out. In either case we will learn something about galaxy formation and
evolution and the dark matter structures of these lowest mass galaxies.

\acknowledgements 
We thank Asa Bluck, Nick Kaiser, Joe Silk, Marta Volonteri, Vicky Kalogera, and Graham Smith for discussions on this topic, comments on this paper, and for their insights.   Support from the Observatoire de Paris where this work was started is gratefully acknowledged.  CJC and RB acknowledge support from a STFC studentship.



\end{document}